\documentclass[10pt,draftcls,onecolumn]{IEEEtran}
\usepackage[utf8]{inputenc}
\usepackage{amsmath,mathtools}
\usepackage{amsthm}
\usepackage{amsfonts}
\usepackage{amssymb}
\usepackage{graphics}
\usepackage{float}
\usepackage{tikz}
\usepackage{caption}
\usepackage{graphicx}
\usepackage{booktabs}
\usepackage[rightbars]{changebar}
\usetikzlibrary{shapes,snakes,patterns}
\usepackage{epstopdf}
\usepackage[normalem]{ulem}
\renewcommand{\vec}[1]{\mathbf{#1}}
\usepackage[justification=centering]{caption}
\usepackage{multirow}
\usepackage{multicol}
\usepackage{enumerate} 
\usepackage{cite}
\usepackage{placeins}
\usepackage{breqn}
\newcommand{\vast}{\bBigg@{4}}
\newcommand{\Vast}{\bBigg@{5}}
\newtheorem{theorem}{Theorem}[]
\newtheorem{lemma}[]{Lemma}
\usepackage{suffix}
\usepackage{mathtools}
\usepackage{amsthm}
\usepackage{enumitem}
\usepackage{xfrac}
\usepackage{relsize}
\usepackage{dsfont}
\usepackage{url}
\usepackage{xcolor}
\usepackage{slashbox}
\theoremstyle{definition}
\usepackage[list=true]{subcaption}
\renewcommand{\th}{\ensuremath{-\text{th } }}

\newcommand{\E}[1]{\mathbb{E}\left[#1\right]}

\newcommand{\xsum}[4]{\sum\limits_{\substack{#1=#2\\ #4}}^{#3}}

\newcommand{\ps}[1]{e^{j#1}}

\newcommand{\mrk}[1]{m_{#1}^{rk}}

\newcommand{\msr}[1]{m_{#1}^{sr}}
\newcommand{\g}{\Vec{g}}
\newcommand{\s}{$S $ }

\newcommand{\G}{\mathbf{G}}

\newcommand{\diag}[1]{\operatorname{diag}\left(#1\right)}

\begin{document}
\bstctlcite{IEEEexample:BSTcontrol}
\title{SINR Analysis of an IRS Assisted MU-MISO System }
\author{Lakshmi Jayalal, Shashank Shekhar, Athira Subhash,  and Sheetal Kalyani \\ 
\thanks{Lakshmi Jayalal, Shashank Shekhar, Athira Subhash, and Sheetal Kalyani are with the Dept. of Electrical Engineering, Indian Institute of Technology, Madras, India. Emails: \{ee19d751@smail,ee17d022@smail,ee16d027@smail, and skalyani@ee \} .iitm.ac.in.}
}
\maketitle
\begin{abstract}
    In this work, we characterize the outage probability (OP) of an intelligent reflecting surface (IRS) assisted multi-user multiple-input-single-output (MU-MISO) communication system. Using a two-step approximation method, we approximate the signal-to-interference-plus-noise ratio (SINR) for any downlink user by a Log-Normal random variable. The impact of various system parameters is studied using the closed-form expression of OP. It is concluded that the position of IRS has a  critical role, but an appropriate increase in the number of IRS elements would help to compensate for the loss in performance if the position of IRS is suboptimal.
\end{abstract}
\section{Introduction}
Intelligence reflecting surface (IRS) has gained much  interest in the wireless communication research community thanks to its promise to enhance the performance of wireless communication systems. IRS is envisioned to be one of the critical enablers of technology for future generations of wireless communication \cite{bariah2020prospective,saad2019vision,akyildiz20206g}. IRS is a hypersurface made up of metamaterials that can interact with impinging electromagnetic (EM) wave and configure it in a programmed manner. A hypersurface planner array consists of several independent meta-atoms \cite{liaskos2018new}. These meta-atoms, also known as IRS elements, can reflect the incident EM wave by configuring the amplitude and phase of the wave \cite{wu2019towards}. 
\par Outage probability (OP) is an important metric to characterize the performance of a communication system. Works like \cite{Char2021Out,tao2020performance,van2021outage} have focused on the characterization of OP for an IRS-assisted single-input-single-output (SISO) systems. IRS-assisted beam index modulation schemes for improved spectral efficiency were proposed in \cite{gopi2020intelligent}. A closed form expression for optimal BF vector is proposed in \cite{subhash2022max} for a single-input-multiple-output system model. Performance analysis of an IRS-assisted multiple-input-single-output (MISO) communication system under a deterministic source to IRS link has been done in \cite{guo2020outage}. The authors of \cite{coelho2020large} considered a single-user MISO system and derived the bit error probability by approximating the effective channel between source and user as a Gamma random variable (RV).  In \cite{Fang2020Outmin}, the stochastic gradient descent method was utilized to jointly design the phase shift at IRS and beamforming (BF) vector at source to minimize the OP where they used the empirical definition of OP as the objective function.  The authors of \cite{zhang2021reconfigurable} consider a multi-user MISO (MU-MISO) system and  provide a lower bound on the number of IRS elements such that the sum rate is greater than a threshold asymptotically in the absence of line-of-sight (LoS) link. Recently, authors of \cite{tota2022Hardware} studied the effect of hardware impairments on an IRS-assisted single-user MISO system where the LoS is assumed to be blocked. They approximated the amplitude of the effective cascaded channel between source and user as Gamma RV using the method of moments.  To the best of our knowledge, all the previous works analyzed the OP of an IRS-assisted MISO system with a single user. 
\par Motivated by this research gap, in this work, we considered an IRS-assisted MU-MISO system. In a multi-user scenario, the signal-to-interference-plus-noise ratio (SINR) is a ratio of two correlated random variables, which makes the statistical characterization of SINR a non-trivial problem. We propose a simple expression for the probability density function (PDF)/ cumulative distribution function (CDF) of SINR using a two-step approximation. The approach developed in this work is quite general and can be easily applied to analyze the OP for a system with other general fading channels.      

\textbf{Notation:}
In this work, $\mathcal{LN}(\mu, \sigma^2)$ represent the Log-Normal distribution with parameters $\mu$ and $\sigma$, $\operatorname{diag}(\theta_{1},\cdots,\theta_{N})$ denotes a diagonal matrix with entries $\theta_{1},\cdots,\theta_{N}$ and $\operatorname{arg(z)}$ denotes the argument (phase) of the complex number $z$. $\operatorname{Cov}(\cdot)$ is the covariance, $\ln$ is the natural log and $\operatorname{erfc}(\cdot)$ is the complementary error function \cite{andrews1998special}
\section{System Model}

\begin{figure}[H]
    \centering
    \includegraphics[scale = 0.7]{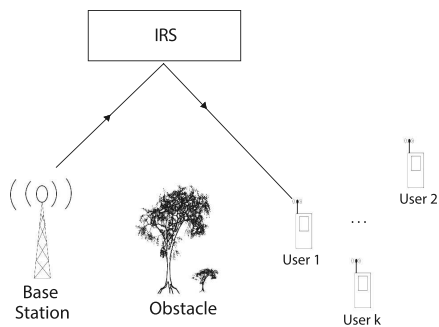}
    \caption{An IRS-Aided multiple input single output (MISO) System}
    \label{fig:Setup}
\end{figure}
We consider an IRS-assisted downlink MISO system similar to the model studied in \cite{zhang2021reconfigurable}. The source, denoted by $\mathbf{S}$, is equipped with $M$ antennas that serve $K$ users, each equipped with a single antenna. The number of IRS elements is assumed to be $N$. The distances from $\mathbf{S}$ to IRS and IRS to $k$-th user ($U_k$)  are represented by $d_{sr}$ and  $d_{rk}$, respectively. We assume that the dimensions of IRS elements are small compared to the distances $d_{rk}$ and $d_{sr}$; hence, all the IRS elements experience the same fast fading. The presence of a blockage between $\mathbf{S}$ and $U_k$ is assumed such that there is no direct link between $\mathbf{S}$ and $U_k$.
\par The channel from $\mathbf{S}$ to IRS and IRS to $U_k$ are assumed to experience independent Rayleigh fading and are represented as $\G_{sr}\in \mathbb{C}^{N\times M}$ and $\vec{h}_{rk}\in \mathbb{C}^{N\times 1}$ ,respectively. The corresponding path losses are given by $d_{sr}^{-\beta}$ and $d_{rk}^{-\beta}$ where $\beta$ is the pathloss coefficient. The phase-shift matrix of IRS is represented as $\Gamma=\alpha\operatorname{diag}({\ps{\theta_1},\cdots,\ps{\theta_N}})$ with reflection coefficient $\alpha =1$ and $\theta_i\in\left[-\pi,\pi\right]$. The effective channel from $\mathbf{S}$ to $U_k$ is represented as $\g_k\in \mathbb{C}^{1\times M}$ where $\g_k=\vec{h}_{rk}^H\Gamma\G_{sr}=\xsum{n}{1}{N}{}h_{rk}^n\alpha\ps{\theta_n}\g_{sr}^n \quad \forall k \in \{1,2,\cdots,K\}$ , $h_{rk}^n$ is the $n$\th element of $\vec{h}_{rk}$, and $\g_{sr}^n$ is the $n$\th row of $\G_{sr}$. 
The transmit symbol is encoded at \s to support multiple users with a precoding matrix $\Vec{w}_k\in\mathbb{C}^{M\times 1}$. Since, for a given IRS phase shift matrix $\Gamma$, maximal ratio transmission (MRT) is the optimal transmit beamforming that maximizes received signal power, similar to \cite{Wu2019Intelli}, we also adopt MRT precoding.
Let the normalized power allocation for user $k$ be $\lambda_k$ such that $\operatorname{Trace}\left(\mathbf{\Lambda\Lambda}^H\right)=1$ where $\mathbf{\Lambda}=\diag{\lambda_1,\cdots,\lambda_K}$.
 Therefore, the received signal at the $U_k$ user can then be written as
\begin{equation}\label{eqn:MUMISOIRS_transmitSignalExpand}
    \begin{aligned}
        y_{k}&= \g_k\g_k^H\lambda_ks_k+\sum\limits_{j\neq k}\g_k\g_j^H\lambda_js_j+n_k,
    \end{aligned}
\end{equation}
where, $\vec{s}=[s_1,\cdots,s_k\cdots,s_K]^T$  is the transmitted signal satisfying $\E{\vec{s}\vec{s}^H}=\frac{P}{K}I_K$ with transmit power $P$.
 For simplicity of the analysis, we are doing uniform power allocation to users, \textit{i.e.,}  $\lambda_k=\frac{1}{\sqrt{K}}$.
Here, $n_k$ represents additive white Gaussian noise at $U_k$ with zero mean and variance, $\sigma_k^2$ . The second term in \eqref{eqn:MUMISOIRS_transmitSignalExpand} corresponds to the interference experienced by the $U_k$ from other users. The received SINR at $U_k$ is given by
 \begin{equation}\label{eqn:MUMISOIRS_SINR}
     \begin{aligned}
      \gamma_k=\frac{X_k}{Y_k}&=\frac{|\g_k\g_k^H\lambda_k|^2}{\sum\limits_{j\neq k}|\g_k\g_j^H\lambda_j|^2+\sigma_k^2}.
     \end{aligned}
\end{equation}
Next, we derive an approximation for the distribution of $\gamma_k$.
 \section{The Log-Normal Approximation} 
 To derive the exact distribution of \eqref{eqn:MUMISOIRS_SINR}, we need to first characterize the joint statistics of $X_k$ and $Y_k$. Here, $X_k$ is an absolute square of the sum of the product of Rayleigh distributed RVs, and $Y_k$ is a sum of the absolute square of the sum of the product of Rayleigh distributed RVs. Note that each RV $X_k$ and $Y_k$ have complicated distributions owing to our choice of MRT BF. Also, the $X_k$ and $Y_k$ are correlated due to  the presence of $\vec{g}_k$ in their expression. Hence, the exact characterization of the distribution of $\gamma_k$ will be a mathematically involved task. Moreover, even if we characterize them in a form that can be evaluated, it will not be easy to use them to derive meaningful inferences or to derive the statistics of other metrics that are a function of SINR. We can circumvent this issue by approximating the distribution of SINR of $U_k$ with a known distribution like Gamma distribution,  log-normal distribution, etc. which is widely used in fitting distributions for positive RVs \cite{Al2010On}. In literature, different variations of correlated bi-variate Gamma distributions \cite{Tubbs1982Bivariate,Chen2014Bivariate,LOAICIGA2005329} are present. Some of these have non-Gamma marginal distributions \cite{van2013Bivariate}, constraints in the correlation coefficient \cite{Izawa1953Bivariate}, and are derived from RVs following other distributions \cite{Piboongungon2005Bivariate}. Deriving the distribution of $\gamma_k$ from correlated bi-variate Gamma distribution results in an expression with infinite sums and integrals. It is very difficult to numerically evaluate these expressions using computer algebra programs like Mathematica. Hence we try to approx the numerator and  denominator by a distribution which leads to an accurate yet simple expression.
  In this work, we first approximate the numerator and the denominator of $\gamma_k$ as Log-Normal RVs. Note that Log-Normal distribution is widely used in fitting distributions for positive RVs \cite{Nair2020}. We then characterize the distribution of $\gamma_k$ as explained in the following paragraphs.
 \begin{lemma}\label{lemma:MUMISOIRS_PDF_X}
 The distribution of $X_{k}$ is approximated as Log-Normal distribution i.e. $X_k\sim \mathcal{LN}(\mu_{X_k},\sigma_{X_k})^2$ with Probability Density Function (PDF) 
\begin{equation}\begin{aligned}
    f_{X_k}(x)&={\frac {1}{x\sigma_{X_k} {\sqrt {2\pi \,}}}}\exp \left(-{\frac {(\ln x-\mu_{X_k} )^{2}}{2\sigma_{X_k} ^{2}}}\right),
\end{aligned}\end{equation}
    where $\mu_{X_k}$ and $\sigma_{X_k}^2$ is obtained as follows:\\
    \begin{equation}\begin{aligned}
        \mu_{X_k}=\ln\left({\frac{\E{X_k}^2}{\sqrt{\E{X_k^2}}}}\right),\quad  \sigma_{X_k}^2=\ln\left({\frac{\E{X_k^2}}{\E{X_k}^2}}\right).
    \end{aligned}\end{equation}
   
 \end{lemma}
 \begin{proof}The approximation is obtained using the method of moments  \cite{Char2021Out}. Here, we evaluate the first and second moments of the RV $X_k$ and solve for the parameters of the Log-Normal distribution with the corresponding values for the moments. Please refer to Appendix \ref{App:MomentMatching_X} for the derivation of the first and second moments of $X_{k}$.
 \end{proof}
\begin{lemma}\label{lemma:MUMISOIRS_PDF_Y}
The distribution of $Y_{k}$  is also approximated as Log-Normal distribution i.e. $Y_k\sim\mathcal{LN}(\mu_{Y_k},\sigma_{Y_k}^2)$ with PDF,
\begin{align}
f_{Y_k}(y)&={\frac {1}{y\sigma_{Y_k} {\sqrt {2\pi \,}}}}\exp \left(-{\frac {(\ln y-\mu_{Y_k} )^{2}}{2\sigma_{Y_k} ^{2}}}\right),
\end{align}
where $\mu_{Y_k}$ and $\sigma_{Y_k}^2$ is obtained as follows:\\
\begin{equation}\begin{aligned}
    \mu_{Y_k}=\ln\left({\frac{\E{Y_k}^2}{\sqrt{\E{Y_k^2}}}}\right),\quad
    \sigma_{Y_k}^2=\ln\left({\frac{\E{Y_k^2}}{\E{Y_k}^2}}\right).
\end{aligned}\end{equation}
\end{lemma}
\begin{proof}
The proof is similar to the proof for Lemma 1. Please refer to Appendix \ref{App:MomentMatching_Y} for the derivation of the first and second moments of $Y_{k}$. 
\end{proof}
Given that $X_k$ and $Y_k$ are now approximated as Log-Normal RVs, by basic transformation of RVs, their ratio \textit{i.e.} $\gamma_k$ will also follow the Log-Normal distribution. Note that the OP can be evaluated as $    P_{\text{out}}(\gamma)=\mathbb{P}(\gamma_{k}\leq \gamma)=F_{\gamma_{\text{IRS}}}(\gamma)$ where $F_X(x)$ is the cumulative density function (CDF) evaluated at $X=x$ and hence, we can use the Log-Normal CDF to characterize the OP for a given threshold. The key result characterizing the OP is given in the following theorem. 
\begin{theorem}
    The OP at $U_k$ for a threshold $\gamma$ is approximated as 
    \begin{equation}\begin{aligned}
        P_{\text{out}}(\gamma)={\frac {1}{2}}\left[\operatorname {erfc} \left(-{\frac {\ln \gamma-\mu_{k} }{\sigma_{k} {\sqrt {2}}}}\right)\right],\label{eqn:MUMISOIRS_poutthm}
    \end{aligned}\end{equation}
    where $\mu_{k} =\mu_{X_k}-\mu_{Y_k}$, $\sigma_k^2=\sigma_{X_k}^2+\sigma_{Y_k}^2-2 \operatorname{Cov}(\ln{X_k},\ln{Y_k})$, and  $\operatorname{Cov}(\ln{X_k},\ln{Y_k})$ can be evaluated using \eqref{eqn:MUMISOIRS_cov}.
\end{theorem}
\begin{proof}
    By the transformation of RVs, the ratio of two correlated Log-Normal distributions is also a Log-Normal distribution \cite{mielke1976distributions}. \textit{i.e.} $\gamma_k=X_k/Y_k$ where $X_k\sim \mathcal{LN}(\mu_{X_k},\sigma_{X_k}^2)$ and $Y_k\sim \mathcal{LN}(\mu_{Y_k},\sigma_{Y_k}^2)$, will be  $\gamma_k\sim \mathcal{LN}(\mu_{\gamma_k},\sigma_{\gamma_k}^2)$ where $\mu_{\gamma_k}=\mu_{X_k}-\mu_{Y_k}$ and $\sigma_{\gamma_k}^2=\sigma_{X_k}^2+\sigma_{Y_k}^2-2\operatorname{Cov}(\ln{X_k},\ln{Y_k})$. Please refer Appendix \ref{App:MUMISOIRS_cov} for the derivation of $\operatorname{Cov}(\ln{X_k},\ln{Y_k})$.
\end{proof}
In the next section, we validate the proposed approximation using the Monte-Carlo simulation.
\section{Simulation Results}
\begin{minipage}[b]{\linewidth}
 \centering
    \begin{tikzpicture}[scale=0.5]
       \draw[->] (0, 0) -- (1, 2) ;
       \draw[->] (1,2) -- (7.8,0.5);
       \draw[dash pattern= on 3pt off 5pt,->] (0,0) -- (2.8,0);
       \draw[gray, very thick] (3,-0.1) rectangle (3.5,0.9);
       \draw[-](7,-0.2)--(7,0);
       \draw[|-|](-0.5,0.1)--(0.5,2.1);
       \draw[|-|](1.2,3) -- (8,1.5);
       \draw [decorate, decoration = {brace,mirror,amplitude=5pt,raise=5pt}] (11,-2.1) --  (11,2);
       \draw [decorate, decoration = {brace,amplitude=5pt,raise=5pt}] (7,2) --  (11,2);
       \draw [decorate, decoration = {brace,mirror,amplitude=5pt,raise=5pt}] (0,-0.1) --  (7,-0.1);
    \draw[dotted,gray, very thick] (7,-2) rectangle (11,2);
      \node at(3.4,-1) {$D$};
      \node at(12,0) {$L$};
      \node at(9,3) {$L$};
      \node at(1,1) {$\G_{sr}$};
      \node at(2,1.2) {$\vec{h}_{rk}$};
      \node at(4,2) {$d_{rk}$};
      \node at (-0.4,1.5) {$d_{sr}$};
      \node at (0, 0) [circle,fill,scale=0.5]{};
     \node at (8, 0.5) [circle,fill,scale=0.5]{};
      \node at (1,2) [circle,fill,scale=0.5]{};
    \node at (-0.5, -1.) []{($0,0$)};
    \node at (-0.5, -0.5) []{$\mathbf{S}$};
    \node at (8, 0.25) []{($x_k,y_k$)};
    \node at (8, 1) []{$U_k$};
    \node at (0.5, 2.5) []{($x_R,y_R$)};
     \node at (.5, 3) []{IRS};
\end{tikzpicture}

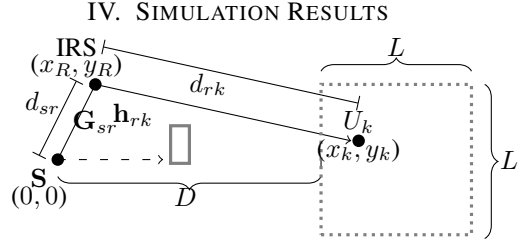
\captionof{figure}{IRS Aided MISO System}
     \label{fig:set_up}
\end{minipage}
In this section, we present the simulation results and verify the  accuracy of the proposed approximation. The layout of the simulation setup is given in Fig. \ref{fig:set_up}. The simulation parameters are adopted from \cite{zhang2021reconfigurable} with $\beta=2$, $M=8$ and $K=10$. The users are uniformly distributed in a square of side length $L=50$ meter (m). The closest side of the square is at $D=150$ m from the source, with its center at $(175,0)$. The IRS is located at $(x_R,y_R) = (0,5)$ unless mentioned otherwise.  The operating carrier frequency is assumed to be $f=5.9$ GHz. The transmit power is set as $P=56$ dBm and the noise power as $\sigma^2 = -96$ dBm.\\
\begin{figure}[H]
    \centering
    \includegraphics[width=0.45\textwidth]{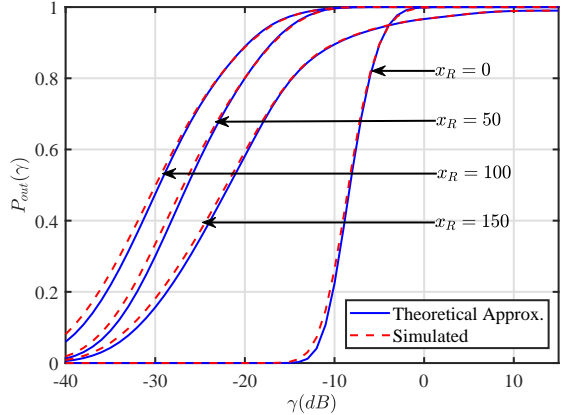}
    \caption{Study of the effect of IRS position ($x_R$) on OP}
    \label{fig:opvsdist}
\end{figure}
In the Fig. \ref{fig:opvsdist}, we study the effect of the position of IRS on OP for $N=50$. In contrast to the single-user system, the OP is not symmetric to the distance of IRS from  $\operatorname{S}$ and $U_k$, which was the case in \cite{Char2021Out}. Due to interference from other users, the OP at a user  depends on interfering users' location and power when the IRS is placed close to the user. Hence, for the best performance, IRS should be placed closer to $\operatorname{S}$, especially in a MU system. 
\begin{figure}[H]
    \centering
    \includegraphics[width=0.45\textwidth]{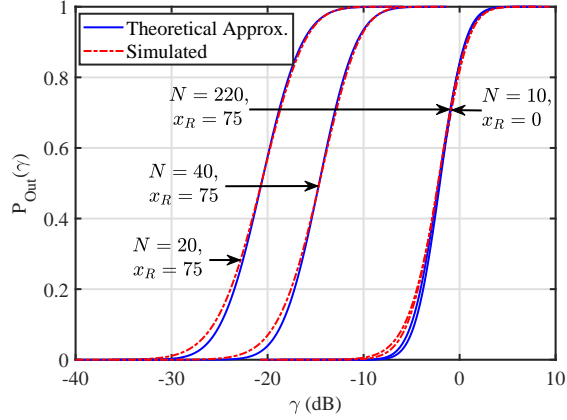}
    \caption{Effect of $N$ on OP}
    \label{fig:opvsdistN}
\end{figure}

In Fig. \ref{fig:opvsdistN}, we study the effect of $N$ on the OP. We can see from the figure that for $x_r=75$, an additional $210$ IRS elements is required to obtain the same OP when $x_r=0$. Thus, it is inferred that the number of IRS elements can be varied to compensate for the loss in performance. We can also observe that the OP at the user decreases as the number of IRS elements increases. Thus, increasing the number of IRS elements can compensate for the increase in OP due to factors like path loss, decrease in transmit power, etc.
\section{Conclusion}
In this work, we have approximated the SINR of an IRS-assisted MU-MISO system by a Log-Normal RV. Based on this, we have characterized the OP in an MU-MISO system without a direct LoS link. Numerical results confirmed the closeness of the proposed approximation to the simulated OP. We also studied the impact of IRS position and the number of IRS elements on the OP. The tractable closed-form approximation provided here can also be used to derive other performance metrics, which are functions of the SINR. Also, the methodology presented in this work is general and can be extended to other fading channel scenarios.
\appendices
\section{Proof of Lemma \ref{lemma:MUMISOIRS_PDF_X}}\label{App:MomentMatching_X}
From \eqref{eqn:MUMISOIRS_SINR}, we have $X_k =\lambda_{k}^2(\g_k\g_k^H )^2$. Let,  $\E{|h_{rk}^j|^i}=\mrk{i}$, $\E{|\g_{sr}^j(l)|^i}=\msr{i}$, $\forall i=\left\{2, 4, 6, 8\right\}$,  $m^{xy}_{2}{}=\alpha_{xy}=d_{xy}^{-\beta}$ where $xy\in\{sr,rk,rj,rh\}$. From the property of moments of Rayleigh distribution, $m^{xy}_{4}{}=2\alpha_{xy}^2$, $m^{xy}_{6}{}=6\alpha_{xy}^3$, $m^{xy}_{8}{}=24\alpha_{xy}^4$, where $xy\in\{sr,rk,rj,rh\}$. After expanding and substituting, the first moment is obtained as,
\begin{equation}\label{eqn:MUMISOIRS_NrE}
    \begin{aligned}
    \E{X_k} &=\lambda_k^2 \alpha_{rk}^2 \alpha_{sr}^2 M N(M+1)(N+1).
\end{aligned}
\end{equation}
Similarly, the second moment is evaluated as 

\begin{equation}\label{eqn:MUMISOIRS_NrE2}
\begin{aligned}
    \E{X_k^2}=\lambda_{k}^4\alpha_{rk}^4 \alpha_{sr}^4 M N (48 \alpha_{sr}^2 (M-2) (M-1) (N-1) +M^3 (N+1) (N+2) (N+3)+6 M^2 (N (N (N+6)\\+3) +14)+M (N (11 N (N+6)+265)-78)+6 (N (N (N+6)-5)+22)).
\end{aligned}
\end{equation}
\section{Proof of Lemma \ref{lemma:MUMISOIRS_PDF_Y}}\label{App:MomentMatching_Y}
From \eqref{eqn:MUMISOIRS_SINR}, we have $Y_k = \underbrace{\sum\limits_{j\neq k}^{K} \lambda_j^2(\g_k\g_j^H\g_j\g_k^H)}_{Z_{k}}+\sigma_k^2.$ Taking the term-by-term expectation, we have  
 \begin{equation}
    \begin{aligned}
         \E{Y_{k}}
        &=
        \E{Z_{k}} + \sigma_k^2\\
         &= M N (M+N)  \alpha_{rk}\alpha_{sr}^2  \left(\sum\limits_{j\neq k}^{K}\lambda_j^2\alpha_{rj}\right)  + \sigma_k^2.
    \end{aligned}
 \end{equation}
The second moment of $Y_k$ is 
\begin{equation}
    \begin{aligned}
         \E{Y_{k}^{2}} &= \E{Z_{k}^{2}} + 2 \E{Z_{k}}\sigma_k^2 + \sigma_k^4   
    \end{aligned}
\end{equation}
The second moment of $Z_k$ is obtained as,
\begin{equation}\label{eqn:MUMISOIRS_DrE2}
    \begin{aligned}
        \E{Z_{k}^{2}}
        &=\sum\limits_{j\neq k}^{K}\lambda_j^4\E{ \underbrace{\left(\g_k\g_j^H\g_j\g_k^H\right)^2}_{T_{1}}}  +\sum\limits_{j\neq k}^{K}\sum\limits_{h\neq k;h\neq j}^{K}\lambda_j^2\lambda_h^2\E{\underbrace{\g_k\g_j^H\g_j\g_k^H\g_k\g_h^H\g_h\g_k^H}_{T_{2}}}
    \end{aligned}
\end{equation}
Expanding each sum and taking the moment of the $T_{1}$, we get
\begin{equation}\label{eqn:MUMISOIRS_ET1}
    \begin{aligned}
        \E{T_{1}} &= 2MN(M+1)(N+1)(M+N+1)\times  (M+N+2) \alpha_{rj}^2 \alpha_{rk}^2 \alpha_{sr}^4
    \end{aligned}
\end{equation}
Similarly, for $T_{2}$, the moment obtained is 
\begin{equation}\label{eqn:MUMISOIRS_ET2}
    \begin{aligned}
        \E{T_{2}} &= M N(M+1)  (N+1) (M+N+1)\times (M+N+2) \alpha_{rh} \alpha_{rj} \alpha_{rk}^2 \alpha_{sr}^4
    \end{aligned}
\end{equation}
After substituting values from \eqref{eqn:MUMISOIRS_ET1} and \eqref{eqn:MUMISOIRS_ET2} in \eqref{eqn:MUMISOIRS_DrE2}, we have 
\begin{equation}
    \begin{aligned}
        \E{Z_{k}^{2}} &= A_{M,N} \alpha_{rk}^{2}\alpha_{sr}^{4}  \left(\sum\limits_{j\neq k}^{K}\lambda_j^4\alpha_{rj}^{2}\right) + B_{M,N} \left( \E{Z_{k}} \right)^{2}
    \end{aligned}
\end{equation}
where $A_{M,N} = M N(M+1)  (N+1) (M+N+1)(M+N+2) $ and $ B_{M,N} = \frac{(M+1)(M+N+1)(M+N+2)}{M N (N+1)}$.

\section{Expression for $\operatorname{Cov}(\ln (X_k),\ln (Y_k))$}\label{App:MUMISOIRS_cov}
The covariance between $\ln (X_k)$  and  $\ln (Y_k)$ under the Log-Normal assumption is \cite{vzerovnik2013transformation}
\begin{equation}\label{eqn:MUMISOIRS_cov}
    \begin{aligned}
        \operatorname{Cov}(\ln (X_k),\ln (Y_k))=\ln{\left(\frac{\operatorname{Cov}(X_k,Y_k)}{\E{X_k}\E{Y_k}} +1 \right)}
    \end{aligned}
\end{equation}
The covariance between $X_k, Y_k$ is defined as $\operatorname{Cov}(X_k,Y_k)=\operatorname{Cov}(X_k,Z_k)=\E{(X_k-\E{X_k})(Y_k-\E{Y_k})}$
\begin{equation}
    \begin{aligned}
        \operatorname{Cov}(X_k,Y_k)=\E{X_kZ_k}-\E{X_k}\E{Z_k},
    \end{aligned}
\end{equation}
where $\E{X_kZ_k}$ is obtained as follows,
\begin{align}
    \E{X_kZ_k}=\alpha_{rk}^3 \alpha_{sr}^4\lambda_k^2 C_{M,N}\sum\limits_{j\neq k}^K\lambda_j^2\alpha_{rk},
\end{align}
where $C_{M,N}=M N (M^3 (1 + N) (2 + N) +    M^2 (1 + N) (2 + N) (5 + N) + M (1 + N) (2 + N) (8 + 3 N) +    2 (N-1) (14 + N (6 + N)))$
\bibliography{ref}
\bibliographystyle{IEEEtran}
\end{document}